\documentstyle[11pt,aaspp]{article}

\begin{document}

%
%
%
   \def\prd#1#2#3#4{#4 19#3 Phys.~Rev.~D,\/ #1, #2 }
   \def\physd#1#2#3#4{#4 19#3 Physica ~D,\/ #1, #2 }
   \def\pl#1#2#3#4{#4 19#3 Phys.~Lett.,\/ #1, #2 }
   \def\prl#1#2#3#4{#4 19#3 Phys.~Rev.~Lett.,\/ #1, #2 }
   \def\pr#1#2#3#4{#4 19#3 Phys.~Rev.,\/ #1, #2 }
   \def\prep#1#2#3#4{#4 19#3 Phys.~Rep.,\/ #1, #2 }
   \def\pfl#1#2#3#4{#4 19#3 Phys.~Fluids,\/ #1, #2 }
   \def\pps#1#2#3#4{#4 19#3 Proc.~Phys.~Soc.,\/ #1, #2 }
   \def\nucl#1#2#3#4{#4 19#3 Nucl.~Phys.,\/ #1, #2 }
   \def\mpl#1#2#3#4{#4 19#3 Mod.~Phys.~Lett.,\/ #1, #2 }
   \def\apj#1#2#3#4{#4 19#3 Ap.~J.,\/ #1, #2 }
   \def\aj#1#2#3#4{#4 19#3 Astr.~J.,\/ #1, #2}
   \def\acta#1#2#3#4{#4 19#3 Acta ~Astr.,\/ #1, #2}
   \def\rev#1#2#3#4{#4 19#3 Rev.~Mod.~Phys.,\/ #1, #2 }
   \def\nuovo#1#2#3#4{#4 19#3 Nuovo~Cimento~C,\/ #1, #2 }
   \def\jetp#1#2#3#4{#4 19#3 Sov.~Phys.~JETP,\/ #1, #2 }
   \def\sovast#1#2#3#4{#4 19#3 Sov.~Ast.~AJ,\/ #1, #2 }
   \def\pasj#1#2#3#4{#4 19#3 Pub.~Ast.~Soc.~Japan,\/ #1, #2 }
   \def\pasp#1#2#3#4{#4 19#3 Pub.~Ast.~Soc.~Pacific,\/ #1, #2 }
   \def\annphy#1#2#3#4{#4 19#3 Ann. Phys. (NY), \/ #1, #2 }
   \def\yad#1#2#3#4{#4 19#3 Yad. Fiz.,\/ #1, #2 }
   \def\sjnp#1#2#3#4{#4 19#3 Sov. J. Nucl. Phys.,\/ #1, #2 }
   \def\astap#1#2#3#4{#4 19#3 Ast. Ap.,\/ #1, #2 }
   \def\anrevaa#1#2#3#4{#4 19#3 Ann. Rev. Astr. Ap.,\/ #1, #2
                       }
   \def\mnras#1#2#3#4{#4 19#3 M.N.R.A.S.,\/ #1, #2 }
   \def\jdphysics#1#2#3#4{#4 19#3 J. de Physique,\/ #1,#2 }
   \def\jqsrt#1#2#3#4{#4 19#3 J. Quant. Spec. Rad. Transfer,\/ #1,#2 }
   \def\jetpl#1#2#3#4{#4 19#3 J.E.T.P. Lett.,\/ #1,#2 }
   \def\apjl#1#2#3#4{#4 19#3 Ap. J. (Letters).,\/ #1,#2 }
   \def\apjs#1#2#3#4{#4 19#3 Ap. J. (Supp.).,\/ #1,#2 }
   \def\apl#1#2#3#4{#4 19#3 Ap. Lett.,\/ #1,#2 }
   \def\astss#1#2#3#4{#4 19#3 Ap. Sp. Sci.,\/ #1,#2 }
   \def\nature#1#2#3#4{#4 19#3 Nature,\/ #1,#2 }
   \def\spscirev#1#2#3#4{#4 19#3 Sp. Sci. Rev.,\/ #1,#2 }
   \def\advspres#1#2#3#4{#4 19#3 Adv. Sp. Res.,\/ #1,#2 }
   %
%
%
\def\Msun{M_{\odot}}
\def\Mdot{\dot M}
\def\deg{$^\circ$\ }
\def\etal{{\it et~al.\ }}
\def\eg{{\it e.g.,\ }}
\def\etc{{\it etc.}}
\def\ie{{\it i.e.,}\ }
\def\ksec{{km~s$^{-1}$}}
\def\arcsec{{$^{\prime\prime}$}}
\def\arcmin{{$^{\prime}$}}
\def\subsun{_{\twelvesy\odot}}
\def\sun{\twelvesy\odot}
\def\gtwid{\mathrel{\raise.3ex\hbox{$>$\kern-.75em\lower1ex\hbox{$\sim$}}}}
\def\ltwid{\mathrel{\raise.3ex\hbox{$<$\kern-.75em\lower1ex\hbox{$\sim$}}}}
\def\plusminus{\mathrel{\raise.3ex\hbox{$+$\kern-.75em\lower1ex\hbox{$-$}}}}
\def\minusplus{\mathrel{\raise.3ex\hbox{$-$\kern-.75em\lower1ex\hbox{$+$}}}}

\title{The Quasi-Stationary Structure of Radiating Shock Waves.\\
 I. The One-Temperature Fluid}

\author{Mark W. Sincell}
\affil{Department of Physics \\ University of Illinois at
Urbana-Champaign \\1110 W. Green Street \\Urbana, IL 61801-3080}
\authoraddr{Department of Physics \\ University of Illinois at
Urbana-Champaign \\1110 W. Green Street \\Urbana, IL 61801-3080}
\author{Michael Gehmeyr and Dimitri Mihalas}
\affil{Department of Astronomy, University of Illinois at Urbana-Champaign
\\Laboratory for Computational Astrophysics, National Center for Supercomputing
Applications}
\authoraddr{Department of Astronomy \\The University of Illinois at
Urbana-Champaign \\1110 W. Green Street \\Urbana, IL 61801-3080}

\begin{abstract}

We calculate the quasi-stationary structure of a radiating shock wave 
propagating
through a spherically symmetric shell of cold gas by solving  
the time-dependent equations of
radiation hydrodynamics on an adaptive grid. 
We show that this code successfully resolves the shock wave in both the 
subcritical and supercritical cases and,
for the first time, we have 
reproduced all the expected features
-- including the optically thin temperature spike
at a supercritical shock front -- without invoking analytic
jump conditions at the discontinuity.
We solve the full moment equations for the radiation flux and energy
density, but 
the shock wave structure can also be reproduced if the radiation flux is 
assumed to
be proportional to the gradient of the energy density (the diffusion
approximation), as long as the radiation energy density is determined by
the appropriate radiative transfer moment equation.  

We find that Zel'dovich and Raizer's analytic solution for the shock wave
structure accurately describes a subcritical shock but it underestimates
the gas temperature, pressure, and the radiation flux in the gas ahead of
a supercritical shock.  We argue that this discrepancy is a consequence of
neglecting terms which are second order in the minimum shock compression
ratio [$\eta_1 = (\gamma-1)/(\gamma+1)$, where $\gamma$ is the adiabatic
index] and the inaccurate treatment of radiative transfer near the
discontinuity.  In addition, we verify that the maximum temperature of the
gas immediately behind the shock is given by $T_{+} = 4 T_1/(\gamma+1)$,
where $T_1$ is the gas temperature far behind the shock. 

\end{abstract}


Correspondence to: M. W. Sincell

\section{Introduction}

The structure and dynamics of radiative shock waves in astrophysical
fluids are difficult to model because processes in the shock front occur
on length scales that are many orders of magnitude smaller than the
typical length scales for other gradients (\eg the velocity field in an
accretion flow) in the fluid variables.  There are two standard methods
for computing the structure of shocked fluids.  The first is to treat the
shock as a discontinuity and invoke conservation laws to relate physical
quantities on either side of the shock.  Analytic models of shock waves
have been constructed using this approach (\eg Zel'dovich and Raizer
1967, Heaslet and Baldwin 1963)  but these solutions require many
simplifying assumptions which limit the applicability of the results. 
Time-dependent analytic solutions can be constructed for only the simplest
of model problems (Zel'dovich and Raizer 1967).  The second method is to
introduce an expression
for the viscosity in the shock. This is most common in numerical
solutions, where an artificial viscosity is introduced to spread the shock
over a few grid points. Numerical solutions can be hampered by the large
gradients encountered at the shock front, which lead to unmanageably small
time steps and numerical instabilities.  To overcome these difficulties,
the magnitude of the artificial viscosity is usually chosen to be many
orders of magnitude larger than the physical viscosity. 

Formulating the numerical problem on an adaptive grid can dramatically
increase the effective resolution of the grid and reduce the spurious
effects of artificial viscosity.  Dorfi and Drury (1987) solved the
one-dimensional hydrodynamic equations on an adaptive grid.  They adopted
a simple grid equation which distributes grid points uniformly along the
arc length of a graph of the solution variables and solved two standard
problems: the shock tube and a spherical blast wave.  In both cases, the
adaptive grid concentrated many grid points at discontinuities in the
flow.  Although artificial viscosity is still needed to spread the
discontinuity over a few grid points, the physical separation of each
point is negligible and the shock front appears infinitely steep.  Gehmeyr
and Mihalas (1994) demonstrated that this same equation can be used
to resolve discontinuities in radiating flows and, in particular, they
performed a preliminary numerical study of radiating shock waves.  We
extend the work of Gehmeyr and Mihalas (1994) in this paper. 

An analytic description of the structure of a simple radiating shock
front
is presented in Zel'dovich and Raizer 1967.  The gas upstream of the shock
is assumed to be cold
($T_2 = 0$, see Fig. \ref{fig: shock cartoon}) and at rest.  A piston
(speed $u_p$) drives a shock wave into the cold gas and the wave
propagates into the unshocked material at a speed $D>u_p$. The structure
of the shock front is steady when viewed in a reference frame moving with
the front and, in this frame, the upstream gas flows into the shock at the
shock speed $D$.  The temperature of the shocked gas rises as the kinetic
energy of the cold gas flowing into the shock is converted into internal
energy of the material flowing out of the shock.  The shocked gas moves
away from the discontinuity at a velocity $u_1$, which equals the speed of
a piston driving the gas.  The gas pressure and density also increase
discontinuously at the shock front.  The material upstream is preheated by
radiation from the downstream shocked material; this raises the
temperature of the inflowing material ahead of the shock and cools the gas
behind the shock.

Zel'dovich and Raizer (1967) found that radiating shocks can be divided
into two categories,
subcritical and supercritical, which have qualitatively different
structures.  The shock wave is subcritical if the gas temperature just
ahead of the shock front ($T_{-}$) is smaller than the final post shock
temperature ($T_1$).  In this case, the absorption of radiation by the
cold gas raises the preshock temperature to $T_{-}$.  The gas immediately
behind the discontinuity is heated to $T_{+}$ by compression in the shock
front and then cools radiatively until reaching the final temperature
$T_1$ downstream from the shock.  Radiation from the shocked gas is
absorbed by inflowing cold gas and the gas temperature ahead of the shock
declines exponentially with increasing optical depth, measured from the
discontinuity. A cartoon of the 
subcritical shock structure is presented
in  Fig. \ref{fig: shock cartoon}a. 

If the shock wave is strong enough, radiation from the shocked gas heats
the material upstream from the shock until $T_{-} = T_1$ and the shock
becomes supercritical. Viscous heating raises the temperature of the
shocked gas from $T_{1}$ to $T_{+}$ just ahead of the discontinuity and
the post-shock gas cools radiatively back to $T_1$.  The length scales for
viscous heating and radiative cooling are much less than a radiation mean
free path and the temperature spike ($T_{+} > T_1$) at the gas pressure
discontinuity (indicated by a vertical dashed line in Fig. \ref{fig:
shock cartoon}b) is optically thin.  The upstream temperature can never
exceed the downstream temperature because that would result in a
discontinuity in the radiation flux and accumulation of radiation energy
density, contradicting the steady state assumption (Zel'dovich 1957).  The
gas temperature ahead of the shock falls slowly because the radiation
field remains in equilibrium with the gas.  The gas temperature starts to
drop exponentially several radiation mean free paths upstream of the shock
because the gas and radiation fall out of equilibrium.  A cartoon of the
supercritical shock structure is presented in Fig.  \ref{fig: shock
cartoon}b. 

In this paper we present time-dependent calculations of the structure of
radiating shock waves for a simple model problem, in which a supersonic
piston drives a shock wave through a spherical shell of cold gas at a
constant density.  This test problem is simple enough that we can compare
directly to the analytic solutions of Zel'dovich and Raizer (1967), both
as a test of the code and
to determine the accuracy of the analytic solution. We describe the TITAN
code in \S\ref{subsec: TITAN} and define the test problem in \S\ref{sec:
formulation} We discuss our results in \S\ref{sec: results} and conclude
in \S\ref{sec: conclusions}  

\section{Formulation of the Problem}
\label{sec: formulation}

\subsection{Equations and Methodology}
\label{subsec: TITAN}

We use the TITAN code (Gehmeyr and Mihalas 1994) to solve the time-dependent 
equations of radiation hydrodynamics on an adaptive grid. 
Gehmeyr and Mihalas (1994) provide a detailed description of 
TITAN so we will only summarize the key features of the code here.
The equations of radiation hydrodynamics in the single fluid approximation
are the continuity equation
\begin{equation}
\label{eq: continuity}
D_t(\rho) + \rho { \partial (r^2 u) \over r^2 \partial r} = 0,
\end{equation}
the gas momentum equation
\begin{equation}
\label{eq: gas momentum}
\rho D_t(u) + {\partial P_g \over \partial r} + {\partial (r^3 P_Q) \over
r \partial r} - {\rho \kappa \over c} F = 0,
\end{equation}
the radiation momentum equation
\begin{equation}
\label{eq: rad momentum}
\rho D_t( {F \over \rho c^2} ) + {\partial P_r \over \partial r}
+ { {3P_r - E_r}\over{r} } + {F \over c^2} {\partial u \over \partial r}
+ {\rho \kappa \over c} F = 0,
\end{equation}
the total (gas plus radiation) energy equation
\begin{equation}
\label{eq: fluid energy}
\rho D_t(e_g + {E_r \over \rho}) + { \partial(r^2 F) \over r^2 \partial r}
+ (P_g + P_r) {\partial(r^2 u) \over r^2 \partial r}
+ P_Q \left[ {\partial u \over \partial r} - {u \over r} \right]
+ (E_r - 3 P_r) {u \over r} = 0
\end{equation}
and the radiation energy equation
\begin{equation}
\label{eq: rad energy}
\rho D_t({E_r \over \rho}) + { \partial(r^2 F) \over r^2 \partial r}
+ P_r {\partial(r^2 u) \over r^2 \partial r} 
+ (E_r - 3 P_r) {u \over r} + \rho \kappa c \left(E_r - a_r T^4 \right)
=0,
\end{equation}
where $D_t(x) = \partial x / \partial t + u \partial x / \partial r$ is the
Lagrangean time derivative operator.
We assume a perfect monatomic gas equation of state with an adiabatic
index of 
$\gamma=5/3$ with
a constant absorptive opacity ($\kappa$).
The radiation pressure ($P_r$) and energy density ($E_r$) are related by a
variable Eddington
factor, $f_E = P_r / E_r$.  The Eddington factors are computed with a
formal 
integration of the time-independent radiative transfer equation (\eg Mihalas
and Mihalas 1984) and updated
during each time-step.
The remaining variables in these equations represent the radius ($r$), the 
gas density ($\rho$), 
gas velocity ($u$), gas pressure ($P_g$), gas energy per unit mass ($e_g$), 
gas temperature
($T$) and radiation flux ($F$).  The physical constants in these
equations, and those that follow, 
are the speed of light ($c$), the radiation constant ($a_r$), the proton
mass ($m_p$), the Stephan-Boltzmann constant ($\sigma$) and the Boltzmann
constant ($k_B$).

We incorporate a tensor artificial viscosity (Tscharnuter and Winkler
1979, Gehmeyr and Mihalas 1994)
\begin{equation}
P_Q = -{4 \over 3} \rho \mu_Q \left[ {\partial u \over \partial r}
- {u \over r} \right],
\end{equation}
where
\begin{equation}
\mu_Q = -C_q^2 l^2 min \left[0, {\partial (r^2 u) \over r^2 \partial r} \right],
\end{equation}
$l \sim 10^{-2} r$ and $C_q$ is a constant of order unity.  

The radiation hydrodynamics equation set (eqs. \ref{eq: continuity}, 
\ref{eq: gas momentum}, \ref{eq: rad momentum}, \ref{eq: fluid energy}
and \ref{eq: rad energy}) is supplemented with the adaptive grid equation
(Gehmeyr and Mihalas 1994, Dorfi and Drury 1987).  This equation
distributes grid points so
that
discontinuities in the flow are resolved.  The simplest examples of grids are
the Eulerian grid, where the length scale is uniformly resolved, and
the Lagrangean grid, where the mass of the gas is uniformly resolved. 

The differential equations are written in finite volume form using the
adaptive mesh transport theorem (Winkler, Norman and Mihalas 1984) and
then
expressed as algebraic finite difference equations on a staggered grid
(Gehmeyr and Mihalas 1994).  
The
difference equations are linearized around the current solution and the solution
at the next time step is calculated with a fully-implicit Newton-Raphson 
iteration scheme.  

\subsection{The Model Problem}
\label{subsec: model problem}

We consider a thin shell of gas extending from $R_i=8.0 \times 10^6$~km to
$R_o=8.7 \times 10^6$~km, such as might be found around a neutron star.
This problem is nearly plane parallel because 
$R_o-R_i \ll R_o$.
Initially, the gas is at rest with constant density
($\rho_o = 7.78 \times 10^{-10}$~g/cm$^{3}$) and a shallow temperature profile
\begin{equation}
T(r) = 10 + 75 { {r - R_i} \over {R_o - R_i} } \mbox{ K}.
\end{equation}
The sound speed in the gas is $c_s \ltwid 1$~km/sec.
The gas has a constant absorptive opacity $\rho \kappa = 3.115 \times 
10^{-10}$~cm$^{-1}$.
Initially, the gas and radiation are in equilibrium and that
there is no net flux of radiation. 

At time $t=0$ a piston at $R_i$ starts outward at a constant velocity, $u_p
> c_s$,
and a shock forms ahead of the piston. The shock propagates outward at a
velocity
\begin{equation}
\label{eq: shock speed}
D = u_p / (1 - \eta_{+})
\end{equation}
where $\eta_{+} = \rho_o / \rho_{+}$ is the shock compression ratio and 
$\rho_{+}$
is the gas density behind the shock.  Note that $\eta_{+} \gtwid \eta_{1}$, 
where
$\eta_{1}$ is the final compression ratio, because some additional compression
can occur as the shocked gas cools from $T_{+}$ to $T_1$. 
After a short time, $t \ll (R_o-R_i)/D$, 
the shock reaches a quasi-stationary state in which the structure of the 
shock front is independent of time when viewed in a frame moving at a velocity 
$D$, \ie with the shock front.

In our simple model, the strength of the shock wave is determined
by the piston speed.
If all the kinetic energy of the gas ahead of the shock is converted into
thermal energy of the gas behind the shock, then the post-shock gas temperature
is
\begin{equation}
\label{eq: T_1}
T_1 \simeq {1 \over 2} {m_p \over k_B} \eta_{1} \left( 1-\eta_{1} \right)^{-2} 
u_p^2 \simeq 27 u_{p,5}^2 \mbox{ K},
\end{equation}
where $u_{p,5}$ is the piston speed in $10^5$~cm/sec, which is on the
order of the sound speed in the gas.

\subsection{Analytic Description of the Radiating Shock Front}
\label{subsec: analytic description}

In this section we review the analytic description of radiating shock
waves.  The reasons for this are twofold.  First, the numerical and
analytic calculations should be consistent in the regions of parameter
space where the assumptions of the analytic description are valid.  This
provides a test of the numerical method.  Second, the numerical model will
be used to extend the analytic results into regions where the analytic
approximations fail.  For both applications, it is important to have a
clear understanding of the assumptions of the analytic solution for the
structure of the radiating shock front.  Unless otherwise noted, the
results in this section are found in Zel'dovich and Raizer (1967). 

Zel'dovich and Raizer (1967) solve for the structure of a radiating shock
wave in two steps.
In the first step, they note that the equations of radiation hydrodynamics
conserve mass, momentum and energy. 
These conservation laws translate into three first integrals, which can be 
written in terms of the gas compression ratio, $\eta$:
\begin{eqnarray}
\label{eq: analytic solution}
P_g(\eta) & = & \rho_o D^2 (1 - \eta) \nonumber \\
T(\eta) & = & { {T_1 \eta (1-\eta)} \over {\eta_1 (1-\eta_1)} } \nonumber \\
F(\eta) & = & { {\rho_o D k_B T_1 (1-\eta) (\eta-\eta_1)}
            \over {m_p \eta_1^2 (1-\eta_1)} },
\end{eqnarray}
where we have assumed $P_g \gg P_r$ and that both $T$ and $P_g$ vanish
as $r \rightarrow
\infty$. 
We adopt the convention that the radiation flux is positive 
outwards and set the mean molecular weight to $m_p/2$, which is appropriate
for a fully ionized hydrogen gas.  
We have also
defined
\begin{equation}
\eta_1 = { {\gamma-1} \over {\gamma+1}},
\end{equation}
the minimum post-shock compression ratio for a gas with an adiabatic index of
$\gamma$.
The conservation laws (eq. \ref{eq: analytic solution}) must be satisfied
by the flow outside of the shock front, where the viscosity is negligible.

The compressional work done on the inflowing gas and the change in the
kinetic energy of that gas cancel to order $\eta_1^2$.  Expanding 
equations \ref{eq: analytic solution} to first order in $\eta_1$, one finds
\begin{equation}
\label{eq: approximate energy conservation}
F \simeq {2\rho_o D k_B \over m_p (\gamma-1)} T
\end{equation}
ahead of the shock.
This relation expresses conservation of energy in the preheated gas.

The shocked gas cools from $T_{+}$ to $T_1$ over a radiation mean free path.
This cooling takes place at nearly constant temperature, which implies that
$T - T_1 \propto \eta - \eta_1$, where $\eta-\eta_1 \ll 1$ in the shocked gas.
Expanding equations \ref{eq: analytic solution} to first order in $\eta-\eta_1$,
one finds the approximate energy conservation relation
\begin{equation}
\label{eq: cooling law}
F \simeq {2\rho_o D k_B \over m_p \eta_1 (3-\gamma)} 
\left( T - T_1 \right).
\end{equation}
In the analytic solution, equations \ref{eq: approximate energy conservation}
and \ref{eq: cooling law} are the only link between the radiation transfer
equations and the hydrodynamics.

The second step is the solution of the radiative transfer equation.
In the case of a subcritical shock wave in a gas with a constant opacity, 
the radiation energy density is much larger than the equilibrium value  and
the radiative transfer equation is approximately independent of the
gas temperature.
This decouples the two steps.
The gas and radiation field are in equilibrium near the discontinuity in
a supercritical shock wave and the full set of radiation hydrodynamical 
equations must be solved simultaneously.
Zel'dovich and Raizer (1967)  invoke the approximate conservation law 
(eq. \ref{eq: approximate energy conservation}) to determine an analytic
solution for the structure of a supercritical shock wave.
In the following paragraphs we describe the analytic solution for 
both cases.

The shocked gas is optically thick and a flux
$F_{o} \simeq \sigma T_1^4$ emerges from the discontinuity.  
This flux cools the shocked gas and is then absorbed by the cold material 
ahead of the shock front.
The post-shock gas temperature is fixed by the approximate energy conservation
relation (eq. \ref{eq: cooling law})
\begin{equation}
\label{eq: t plus}
\sigma T_1^4 \simeq {2\rho_o D k_B \over m_p \eta_1 (3-\gamma)} 
\left( T - T_1 \right).
\end{equation}
Similarly, applying the upstream conservation relation 
(eq. \ref{eq: approximate energy conservation}) at the location
of the shock gives the preshock temperature
\begin{equation}
\label{eq: t minus}
\sigma T_1^4 = {2\rho_o D k_B \over m_p (\gamma-1)}
T_{-}.
\end{equation}
The shock speed $D \propto T_1^{1/2}$, so $T_{-} \propto T_1^{7/2}$.
Equating equations \ref{eq: t minus} and 
\ref{eq: t plus} at the shock front, we find
\begin{equation}
\label{eq: shock dt}
T_{+} - T_{-} = T_1 - 2\eta_1 T_{-}
\end{equation}
for the temperature difference across the shock.

The subcritical radiation energy density is greater than the 
equilibrium 
value, $E_{r,eq} = a_rT^4$. 
Neglecting $E_{r,eq}$ in the radiation transfer
equation, one finds
\begin{equation}
\label{eq: subcritical flux profile}
F(\tau) = F_{o} \exp^{-\sqrt{3} |\tau|},
\end{equation}
where $\tau$ is the optical depth of the gas measured from the shock front.
The factor of $\sqrt{3}$ is a consequence of the Eddington approximation and
will change if the radiation field is anisotropic.
As a consequence of 
equations \ref{eq: approximate energy conservation}
and \ref{eq: cooling law}, the gas temperature falls exponentially with
increasing optical depth from the shock front. 

The structure of a supercritical shock is both qualitatively and quantitatively
different from a subcritical shock.  Preheating by the radiant flux raises
$T_{-}$ to $T_1$ when the post shock gas temperature reaches a critical value
\begin{equation}
\label{eq: t critical}
T_c = \left\{ {2 \rho_o D k_B \over \sigma m_p (\gamma-1) } \right\}^{1/3}
\end{equation}
estimated by setting $T_1 = T_{-} = T_c$ in equation \ref{eq: t minus}.  
For the parameters of our model problem, we calculate $T_c \simeq 1600$~K.
When $T_1 \gtwid T_c$, the radiant
energy flux in the preheated gas is comparable to the hydrodynamic flux
and the radiation energy density reaches its equilibrium value.
Two new features appear in the shock structure: an optically thin temperature
spike at the surface of discontinuity and an equilibrium layer in the preheated
gas (see Fig. \ref{fig: shock cartoon}b).

Radiation diffuses through the equilibrium layer.  
Assuming that the approximate energy conservation law 
(eq. \ref{eq: approximate energy conservation})
fixes the relation between $F$ and $T$, Zel'dovich and Raizer (1967) show
that the
radiation flux profiles in this layer are
\begin{equation}
\label{eq: t and f equilibrium}
{T \over T_c} = {\sqrt{3} F \over 4\sigma T_c^4} = \left( 1 + {3\sqrt{3}
\over 4} \left| \tau - \tau_c \right| \right)^{1/3}.
\end{equation}
The radiation field falls out of equilibrium ($T < T_c$) at
\begin{equation}
\label{eq: tau equilibrium}
\left| \tau_c \right| = {4 \over{ 3\sqrt{3} } } \left[ \left( {T_{1} \over T_c}
\right)^{3} - 1 \right].
\end{equation}
and the non-equilibrium solution applies at larger $|\tau|$.
The flux at the discontinuity 
\begin{equation}
\label{eq: fo supercritical}
F_{o} = {4\sigma T_c^4 \over \sqrt{3} } \left( {T_1 \over T_c} \right)
\end{equation}
is found from equation 
\ref{eq: t and f equilibrium}. 

Compression in the shock wave raises the gas temperature at the discontinuity
to $T_{+}$
and radiative cooling rapidly reduces the temperature to $T_1$, resulting in an
optically thin
temperature spike at the discontinuity.  The amplitude of the
spike,
\begin{equation}
\label{eq: t spike}
T_{+} = {4 \over (\gamma+1)} T_{1},
\end{equation}
can be estimated by setting $T_{-} = T_{1}$ in equation
\ref{eq: t plus}.
Raizer (1957) has also derived this relation, but he calls it an
approximation to the exact value of $T_{+} = (3-\gamma) T_1$.  However, we find
that equation \ref{eq: t spike} agrees with the numerical result to within
1\%, whereas the Raizer (1957) expression is only accurate to within 10\%.
The optical depth of the spike 
\begin{equation}
\label{eq: tau spike}
\Delta \tau \sim \left( {T_c \over T_1} \right)^3 
\end{equation}
is determined by equating the flux from the spike 
\begin{equation}
F_s \sim {\sigma T_{1}^4 \over \Delta\tau}
\end{equation}
to the flux at the discontinuity (eq. \ref{eq: fo supercritical}).
The spike is optically
thin ($T_1 > T_c$ by definition) and the thickness decreases rapidly with
increasing shock strength.

\section{Numerical Solution of the Model Problem}
\label{sec: results}

The numerical solution differs from the analytic solution in two respects.
First, we solve the full set of radiation hydrodynamical equations for the
quasi-stationary structure of the shock front.  Second, we assume that
$\gamma = 5/3$, or $\eta_1 = 0.25$.  The analytic solution uses the fact
that $\eta_1 << 1$ to derive a relation between the gas temperature and
the radiation flux without explicitly solving the radiative transfer
moment equations.  This approximation is not as accurate when $\eta_1 = 0.25$.
In the following sections we compare the numerical and analytic results and
discuss the effects of these two differences.
In order to illustrate all the features of the shock structure, we plot
the solution against two axes: $\eta$, the compression
ratio, which emphasizes the region near the discontinuity, and $\tau$, the
optical depth measured from the discontinuity, which emphasizes the heating
and cooling regions near the shock front.

We reemphasize at this point that the numerical solution is a complete
solution of the equations of radiation hydrodynamics for this model
problem (see \S \ref{subsec: model problem}).  The analytic solution
(eq. \ref{eq: analytic solution}) only
provides an exact relation between the gas variables at the end points,
where the radiation flux and viscosity are negligible.  The structure of
actual shock transitions can depart significantly from these curves. 

\subsection{The Subcritical Shock Wave}
\label{subsec: sub}

The numerical results for the temperature and flux in a subcritical shock wave
($u_p = 6$~km/sec) are plotted as a function of the compression ratio 
(Fig. \ref{fig: temp and flux vs eta}) and optical depth from the shock
front
(Fig. \ref{fig: flux vs tau}).  
In these plots, we have scaled the 
temperatures to
$T_1$ and the flux to $F_o = \sigma T_1^4$.  The gas pressure is plotted
as a function of the compression ratio in Fig. \ref{fig: pressure vs eta}.
In each plot, we have superposed the appropriate analytic solution (see
\S\ref{subsec: analytic description}).

The adaptive 
grid has concentrated a large number of points in the shock front and the
structure of the front is well resolved.  
Each point in Figs. \ref{fig: temp and flux vs eta}
and \ref{fig: pressure vs eta} represents a single fluid element.  
The effective resolution
near the shock front exceeds $10^5$, \ie by incorporating adaptive grid with
$\sim 100$ points
we reach a resolution which is equivalent to a fixed grid with $10^5$ points.

The analytic and numerical solutions agree very well in both the shocked 
material
($\eta \ltwid \eta_{+} =
0.3$) and in the preheated gas ($\eta \gtwid \eta_{-} = 0.95$).
The analytic solution correctly links the initial and final states of the
gas, but it slightly underestimates $F$ (Fig. \ref{fig: temp and flux vs
eta}). 
The gas pressure is a linear function of $\eta$ outside of the shock front,
as predicted by eq. \ref{eq: analytic solution}, and each fluid element follows
an approximate Hugoniot curve through the shock front 
(Fig. \ref{fig: pressure vs eta}).  The radiation flux is constant through
the shock front and viscous heating in the front raises the gas temperature.
The maximum compression ratio observed in the shock
is $\eta_{+} \simeq 1/3.6$, so we have not quite reached the strong shock limit
of $\eta_{+} = (\gamma - 1) / (\gamma + 1) = 1/4$.

Although compressional heating is about 50\% larger than the change in the
kinetic energy of the inflowing gas, because terms which are second order 
in $\eta_1$ are not negligibly small, this has a small effect on the structure 
of the shock wave.
There are three, not entirely independent, reasons for this.  First, 
compressional heating of the gas does not change the temperature of the gas
by a large amount. In Fig. \ref{fig: flux vs tau}a, 
it is clear that $T$ declines
nearly exponentially with optical depth from the shock, as expected for 
radiative cooling.   Second, radiative heating and cooling change the gas 
temperature near the shock by $\ltwid 10\%$ of $T_1$.  This minimizes the
consequences of the assumed $F(T)$ relation 
(eq. \ref{eq: approximate energy conservation}).
Third, the radiative transfer equation is independent of the gas temperature
because the radiation energy density is much larger than the equilibrium
value of $a_r T^4$.

When the radiation energy density $E_r \gg a_r T^4$, the radiation flux
falls
exponentially with increasing optical depth from the shock discontinuity.
The numerical solution for $F$, scaled to the flux at the discontinuity,
is plotted with the analytic
approximation in Fig. \ref{fig: flux vs tau}b. 
The agreement between the two
solutions is quite good for $|\tau| \ltwid 1$, where the radiation energy 
density is much larger than the equilibrium value.
At larger $\tau$, the
radiation energy density approaches the equilibrium value and no longer
falls exponentially.  In this region, we find that
the analytic estimate 
underestimates the true flux.

The energy conservation relations 
(eqs. \ref{eq: approximate energy conservation} and \ref{eq: cooling law})
imply that the gas temperature also declines exponentially with optical
depth.  We find that this is largely correct, but a small amount of 
compressional heating raises $T_{-}$ so that the temperature in the preheated
gas falls somewhat faster (Fig. \ref{fig: flux vs tau}a).

\subsection{The Supercritical Shock Wave}
\label{subsec: super}

The shock wave becomes supercritical 
when $u_p \simeq 9-10$~km/sec.
The analytic estimate (eqs. \ref{eq: T_1} and \ref{eq: t critical}) predicts
that $T_1 = T_c$ when $u_p \simeq 7$~km/sec,  
which
is reasonably close to the numerical value.
In the following,
we use a run with $u_p = 16$~km/sec to illustrate the structure of the 
supercritical shock wave.  The structure of the shock front is illustrated
with plots of the physical variables as functions of $\eta$ (Figs. 
\ref{fig: supereta} and \ref{fig: superpeta}) and $\tau$ (Fig. \ref{fig:
super three zone}).

The conservation relations (eq. \ref{eq: analytic solution}) accurately
link the initial and final states of the gas (Fig. \ref{fig: supereta}).
The numerical solution reproduces the isothermal profile of the shocked
gas and the linear dependence of $P$ on $\eta$ downstream from the
discontinuity.  No net energy leaves the system, so the temperature of the
shocked gas is given by eq. \ref{eq: T_1}. Although the agreement between
the numerical and analytic solutions is quite good downstream, the two
solutions differ appreciably in the preheated gas.

We find that the gas temperature, pressure and radiation flux in
the preheated gas is larger than predicted by the analytic solution of
Zeldovich and Raizer (1967).  There are two reasons for the discrepancy.
First, the terms
proportional to $\eta_1^2$  are not vanishingly small.   Physically, this
means that the compressional work done on the inflow is larger than the
change in the kinetic energy of the gas.  This results in more heating of
the gas.  Second, the analytic solution underestimates the radiation flux
in the preheated gas (see Figs. \ref{fig: supereta} and
\ref{fig: super three zone}), particularly within a few radiation mean free
paths of the optically thin temperature spike.  The additional radiation
flux contributes to the local heating rate.  Both effects tend to raise the gas
temperature.  When $\gamma = 4/3$, and the second order terms are smaller,
we find that the compressional work on the gas is very nearly equal to the
change in the kinetic energy and the analytic solution for the gas temperature
is much closer to the numerical result.  However, the radiation flux remains
larger than the analytic prediction.  We attribute this to the additional flux
from the optically thin spike.

The temperature profile of the optically thin temperature spike is
determined by the balance of viscous heating and radiative cooling. In
Fig. \ref{fig: supereta}, the region between $\eta = 0.95$ and $\eta=0.2$
corresponds to the optically thin temperature spike. Viscous heating
raises the gas temperature from $\eta \simeq 0.95$ to $\eta \simeq 0.3$
and the gas cools radiatively at smaller $\eta$ (Fig. \ref{fig: 
supereta}). We use $\gamma = 5/3$ in these simulations, so eq.  \ref{eq: t
spike} predicts $T_{+} = 1.5 T_1$, which is quite close to the numerical
value of $T_{+}= 1.44 T_1$.  The Zel'dovich and Raizer (1967) expression,
$T_{+} = (3-\gamma) 
T_1$, predicts $T_{+} = 1.33 T_1$, which is significantly less accurate
than eq. \ref{eq: t spike}.  The two expressions agree within a few
percent when $\gamma = 4/3$ so the distinction between the two expressions
is unimportant in that case. 

The area under the optically thin temperature spike must be conserved, so
the height of the spike is determined by the (artificial) viscous heating
length and
the radiative cooling length.  In Fig. \ref{fig: super three zone}b, we
have plotted a detailed example of the temperature spike.  It is clear
that the width of the spike is determined by the radiative cooling
length, which implies that the height of the spike does not depend upon
the
coefficient of artificial viscosity. However, we were not able to verify
that the height of the spike reaches a maximum (set by eq. \ref{eq: t
spike}) because we could
not reduce the viscosity to an arbitrarily small value.

Most of the radiation flux is formed in the temperature spike.  The flux
increases from the downstream of the spike to its peak value, $F_o \simeq
{\sigma T_1^4 \over \Delta \tau}$ at the downstream side of the
discontinuity.  The optical depth of the spike is $\Delta\tau$.
The large negative temperature gradient on the
downstream side of the
shock front generates a small flux into (\ie $F < 0$) the shocked gas.
This is not seen in the analytic solution, which assumes that the flux is
continuous across the discontinuity.

The structure of the supercritical shock is more evident when plotted as a
function of optical depth from the shock (Fig. \ref{fig: super three
zone}).  We find that the numerical solution exhibits the expected three
zone structure.  In Fig. \ref{fig: super three zone}a, we plot $T$ and the
radiation temperature $T_r = (E_r / a_r)^{1/4}$ as functions of the
optical depth from the discontinuity. The gas and radiation field are in
equilibrium ($T_r = T$) from the
shock surface to $\tau = \tau_c \simeq -5 $.  The solid line in Fig.
\ref{fig: super three zone} represents the analytic solution for the
equilibrium layer (eq. \ref{eq: t and f equilibrium})  with $T_c = 1700$~K
and $\tau_c = -5$.  Both the temperature profile and the value of $T_c$
are remarkably close to the estimates in \S\ref{subsec: analytic
description}. At larger
$|\tau|$, the radiation drops out of equilibrium and both the radiation
flux and the gas temperature start to drop exponentially with optical
depth. 

The predicted optical depth of the temperature spike is $\Delta\tau =
0.15$.  In Fig. \ref{fig: super three zone}, we plot the temperature over
a
smaller range of $\tau$ to illustrate the structure of the spike. The gas 
and radiation temperatures are out of equilibrium in the spike and we
define the optical depth of the spike as range of $\tau$ for which $T_r >
T$. We find that $\Delta\tau \simeq 0.1$, which is reasonably close to the
analytic estimate, particularly when one considers the many approximations
that are involved in making that estimate. 

In Fig. \ref{fig: super three zone}c), we plot the numerically
determined radiation flux, scaled to the estimated peak flux ($F_o$).
It is clear that the analytic estimate is about 30\% smaller than the
actual peak flux.  The approximate energy
conservation relation (eq. \ref{eq: approximate energy conservation})
implies that $F \propto T$ in the preheated gas, and we find that this is
very nearly satisfied.  The exponential fall-off at $\tau < -5$ is also
verified. 

\subsection{Angular Distribution of the Radiation}
\label{subsec: angular distribution}

Radiation from the subcritical shock is isotropic downstream from the
shock ($f_E \sim 1/3$) and rapidly approaches the free streaming value of
$f_E = 1.0$ upstream from the shock.  The Eddington factors dip below 1/3
at the shock front, indicating that the radiation is preferentially
emitted parallel to the shock front.  This is because radiant energy
generation occurs primarily in a thin shell, approximately a radiation
mean free path across, at the shock radius (eq.
\ref{eq: subcritical flux profile}).  Rays which are nearly parallel to
the shock front have a larger path length for energy generation and the
flux along these directions is correspondingly enhanced.   

The radiation from the supercritical shock remains isotropic from the
piston to the outer edge of the equilibrium region (Fig. \ref{fig: super
eddfac}).  The Eddington factors again fall below 1/3 at the shock radius,
for the reasons just discussed, but in the supercritical case the range of
radii for which $f_E < 1/3$ is smaller.  This is because the temperature
spike at the shock radius is much thinner, $\Delta\tau \ll 1$, than in the
subcritical shock (eq. \ref{eq: tau spike}).  At large radii, the gas
becomes optically thin and the Eddington factors approach the free
streaming value of 1. 

\subsection{Effects of the Assumed Radiation Transport Mechanism}
\label{subsec: transport mechanisms}

The diffusion approximation (\eg Mihalas and Mihalas 1978) is often used
to
simplify the radiative transfer moment equations (eqs. \ref{eq: rad
momentum} and \ref{eq: rad energy}).  In place of the radiation momentum
equation (eq. \ref{eq: rad momentum}), one assumes that the radiation flux
is proportional to the gradient of the radiation energy density
\begin{equation} 
\label{eq: rad diffusion} F = - {c \over 3\rho\kappa}
{\partial E_r \over \partial r}.  
\end{equation} 
This is equivalent to
assuming that the gas is optically thick and isotropic, \ie $f=1/3$. If 
the diffusion approximation is invoked, the equation for the radiation
momentum (eq. \ref{eq: rad diffusion}) is no longer explicitly
time dependent. 

There are two variants of this approximation -- equilibrium and
non-equilibrium diffusion. The gas and radiation temperatures are assumed
to be the same in equilibrium diffusion.  This is a reasonable
approximation if the time and length scales for energy exchange between
the gas and the radiation field are shorter than any scale in the flow. 
In the non-equilibrium diffusion approximation, the radiation energy
density, and $T_r$, is determined by the radiation energy equation (eq.
\ref{eq: rad energy}). 

We find that non-equilibrium diffusion is a good approximation to the full
radiation transport scheme (Fig. \ref{fig: transport comparison}).  The
gas is optically thick over most of the flow and the radiation is nearly
isotropic in all cases (see Figs. \ref{fig: sub eddfac} and \ref{fig:
super eddfac}).  The largest departures from isotropy occur at the shock
front (\S\ref{subsec: angular distribution}) but even there they are
fairly small.  The smallest value of the Eddington factor in the shock
front is $\sim 0.3$. 

Equilibrium diffusion is a reasonable approximation outside of the shock
front but it breaks down inside the front (Fig. \ref{fig: transport
comparison}).  The temperature gradient is very large in the shock and the
temperature changes on scales much shorter than a photon mean free path. 
Thus, the radiation is out of equilibrium in the front (see Fig.
\ref{fig: super three zone}b). Diffusion of
radiation energy, caused by the assumption $T_r = T$, artificially
equalizes the gas temperature on either side of the shock ($T_{+} =
T_{-}$) and the solution follows a path of constant temperature through
the shock front.  In the subcritical shock, this reduces $T_{+}$ and
raises $T_{-}$, relative to the full transport solution.  Assuming
equilibrium diffusion can cause one to overestimate $T_{-}$ by a factor of
3-4 (Fig. \ref{fig: transport comparison}).  In addition, $F$ in the shock
is overestimated by a similar factor because the flux is assumed to be
proportional to the temperature gradient.  Equilibrium diffusion gives an
accurate estimate of $T_{+,-}$ for a supercritical shock, but the extra
energy diffusion erases the most distinctive feature of the shock, the
temperature spike. 

\subsection{Relations between Upstream and Downstream Quantities}
\label{subsec: variable relations}

The post-shock gas temperature ($T_{+}$) increases nearly linearly with
increasing $T_1$.  For a subcritical shock, the
temperature of the shocked gas is related to the temperature of the
preheated gas by eq. \ref{eq: t plus}, and 
\begin{equation} 
T_{+} = T_1 + {1 \over 2} T_{-} 
\end{equation} 
in a $\gamma=5/3$ gas.  We find that this
solution is a good approximation to the numerical solution at $T_{1}
\ltwid 1800$.  Above this temperature the shock becomes supercritical and
$T_{+} \simeq 1.5 T_1$, as predicted by eq. \ref{eq: t spike}.  In Figure
\ref{fig: tp vs t1}, we have plotted the numerical results for several
runs of increasing shock strength (crosses), the relation $T_{+} = 1.5
T_1$ (short dashed line) and the Zel'dovich and Raizer (1967) relation
$T_{+} = 1.33 T_1$. The
two analytic expressions apply to a supercritical shock and the Zel'dovich
and Raizer (1967)
relation underestimates $T_{+}$ by about 10\% in all cases. We find that
eq. \ref{eq: t spike} is a much better description of the numerical
solution. 

Preheating raises the gas temperature ahead of the shock.   We find that
$T_{-} \propto T_1^{7/2}$ (the dashed line in  Fig. \ref{fig: tm vs t1})
in a 
subcritical shock wave, as discused in \S\ref{subsec: analytic description}.
At supercritical shock strengths $T_{-} = T_1$ within a few percent.  The 
temperature at which the shock makes the transition from subcritical to
supercritical ($T_c$) can be estimated as the value of $T_1$ where the
two approximate $T_{-}(T_1)$ relations intersect.  We find that $T_c \simeq
1200$, in reasonable agreement with the estimates in 
\S\ref{subsec: analytic description} and \S\ref{subsec: super}.

\section{Conclusions}
\label{sec: conclusions}

We have solved the time-dependent spherically symmetric equations of
radiation hydrodynamics on an adaptive grid for both subcritical and
supercritical radiating shock waves.  In this paper, the shock wave is
created by a supersonic piston moving into a shell of cold gas with a
constant density. The gas opacity was assumed to be constant and purely
absorptive. The shock propagates into the gas at a speed $D = u_p
/(1-\eta_{+})$, where $\eta_{+}$ is the compression ratio just downstream
of the shock front, and we find that the structure of the shock becomes
steady in the shock frame after a time short compared to the flow time. 
The temperature of the shocked gas is approximately $T_1 \simeq 27
u_{p,5}^2$~K, where $u_{p,5}$ is the piston speed in $10^5$~cm/sec.

The addition of the adaptive grid equation enables the code to fully
resolve the radiating shock, so no jump conditions need to be employed and
the effects of artificial viscosity are greatly reduced.  The code
achieved a peak resolution in the shock front of $3 \times 10^5$ with 100
grid points.  Thus, the grid resolves physical processes occurring on
length scales over $10^3$ times smaller than the best fixed grid scheme. 

The shock wave is subcritical when $u_p \ltwid 9$~km/sec.  The Zel'dovich
and Raizer (1967)
analytic solution for the gas temperature and radiation flux as a function
of the compression ratio agrees with the numerical solution outside of the
shock front.  A single fluid element passes through the shock on an
approximate Hugoniot curve and the radiation flux is constant, and the gas
temperature increases, along this curve. 

The radiation flux from a subcritical shock wave peaks at the shock radius
and declines in proportion to $\exp^{-\sqrt{3} |\tau|}$, where $\tau$ is
the optical depth of the gas measured relative to the shock front, when
$|\tau| \ltwid 1$.  At larger optical depths the radiation energy density
approaches its equilibrium value and the actual flux falls less rapidly
than the analytic model predicts.  The Eddington factors are less than 1/3
at the shock front, indicating that the radiation is directed parallel to
the shock. 

The numerical solution for the supercritical shock wave ($u_p \gtwid
9$~km/sec) exhibits the expected three zone structure: shock heated gas at
$T=T_1$, an optically thin temperature spike at the gas pressure
discontinuity and preheated gas in equilibrium with the radiation field
(Zel'dovich and Raizer 1967).  We find that the amplitude of the
temperature spike is $T_{+} =
4T_1 / (\gamma+1)$.  The radiative cooling length in the spike is much
shorter than the scale of the flow, so the adaptive grid must be used to
resolve the spike.  Fixed grid methods and equilibrium
radiation diffusion introduce too much energy diffusion and the spike
disappears from the numerical solution.  Heating by the radiation from the
shock heats gas ahead of the shock to $T_{-} = T_1$.  The radiation energy
density remains near its equilibrium value, and $T_r = T$, until $\tau
\gtwid -5$, where $\tau$ is measured relative to the shock front.  In this
region, the gas temperature is roughly $\propto |\tau|^{1/3}$, in
agreement with the analytic approximation (Zel'dovich and Raizer 1967).

Although the jump conditions correctly link the initial and final states
of the gas, the analytic solution underestimates the gas temperature,
pressure and the radiative flux.  Optically thin emission from the spike,
not included in the analytic solution, increases the radiation flux near
the shock.  In addition, the analytic solution assumes that the
compressional work done on the flow is exactly equal to the change in the
kinetic energy of the gas.  We find that the compression work is typically
larger than the change in kinetic energy and the discrepancy increases
with $\eta_1$.  Both effects raise the temperature of the preheated gas.

\vfil\eject
\centerline{\bf Figure Captions}
\bigskip

\begin{figure}
\plotone{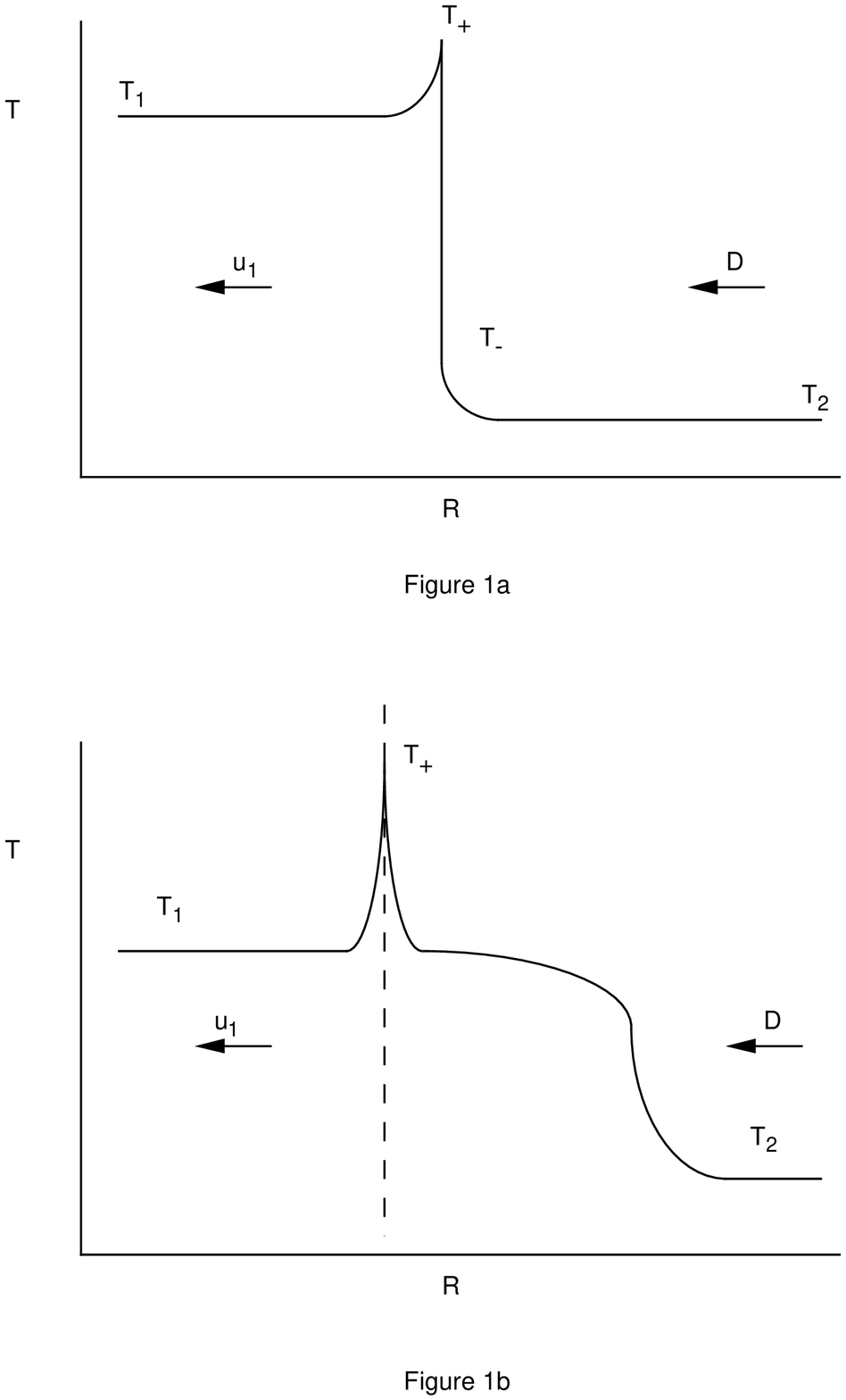}
\caption{\label{fig: shock cartoon}   }
\end{figure}

Figure \ref{fig: shock cartoon}.  The structure of a subcritical (a)
and a supercritical (b)
radiating shock wave.

\begin{figure}
\plotone{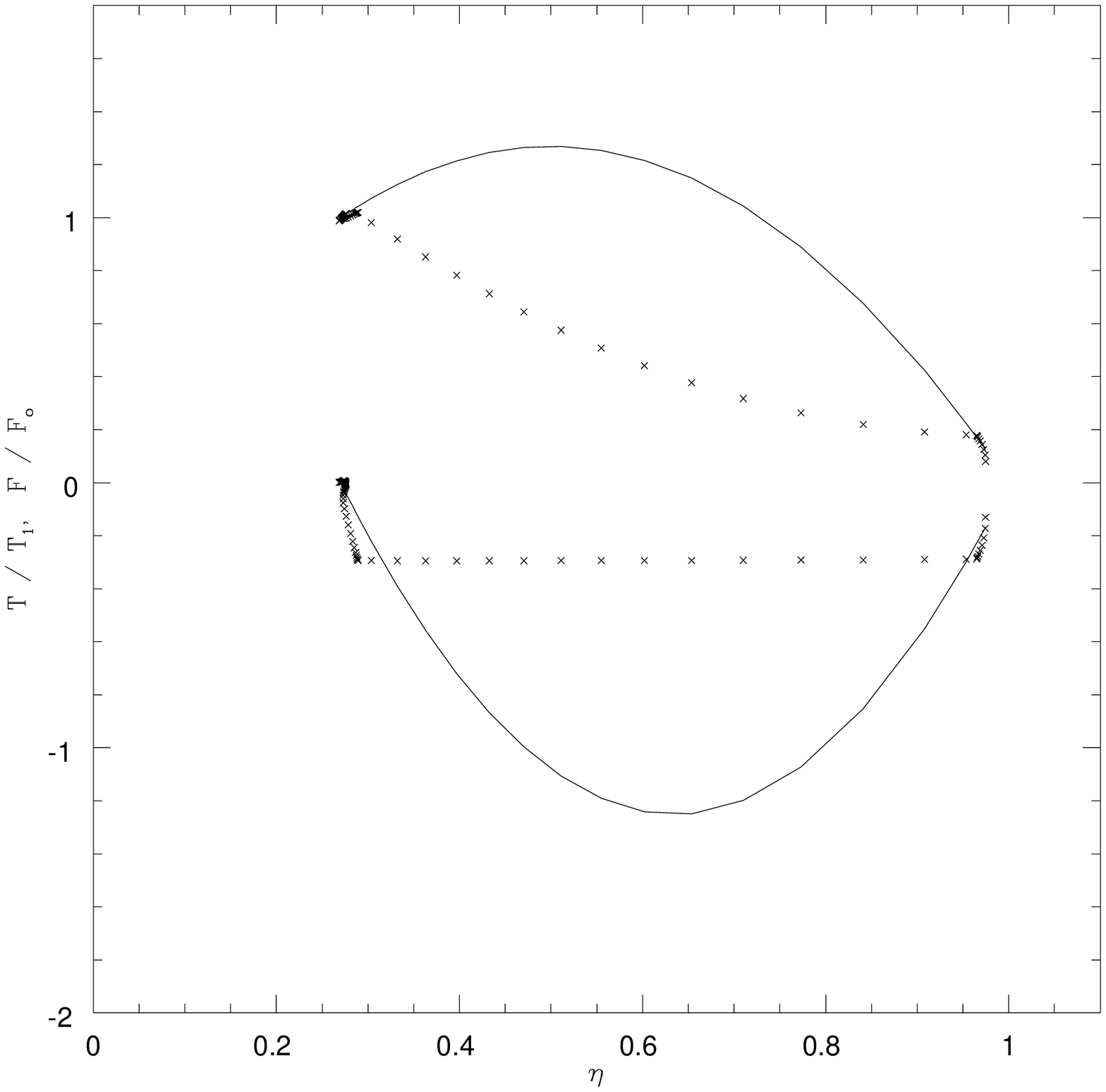}
\caption{\label{fig: temp and flux vs eta}   }
\end{figure}

Figure \ref{fig: temp and flux vs eta}. Gas temperature (upper curve) and
radiation flux (lower curve)
as a function of compression ratio ($\eta$).  The solid curves are the
analytic
expressions from ZR and the crosses are the numerical solution.  
Note that in this figure we plot (-$F$).

\begin{figure}
\plotone{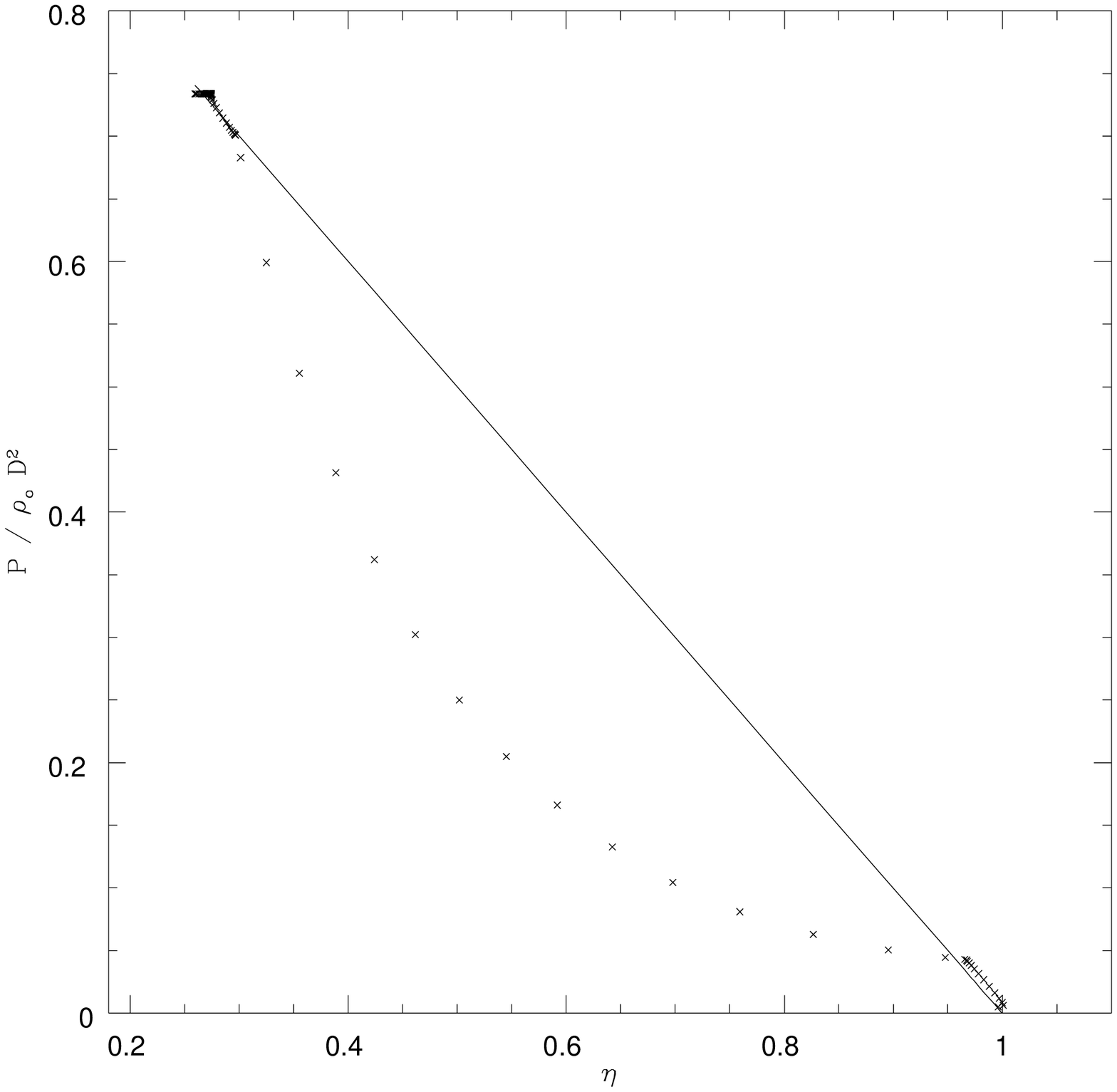}
\caption{\label{fig: pressure vs eta}   }
\end{figure}

Figure \ref{fig: pressure vs eta}. Gas pressure as a function of compression
ratio.  The solid line is the analytic expression from ZR and the dashed line
is the numerical result.

\begin{figure}
\plotone{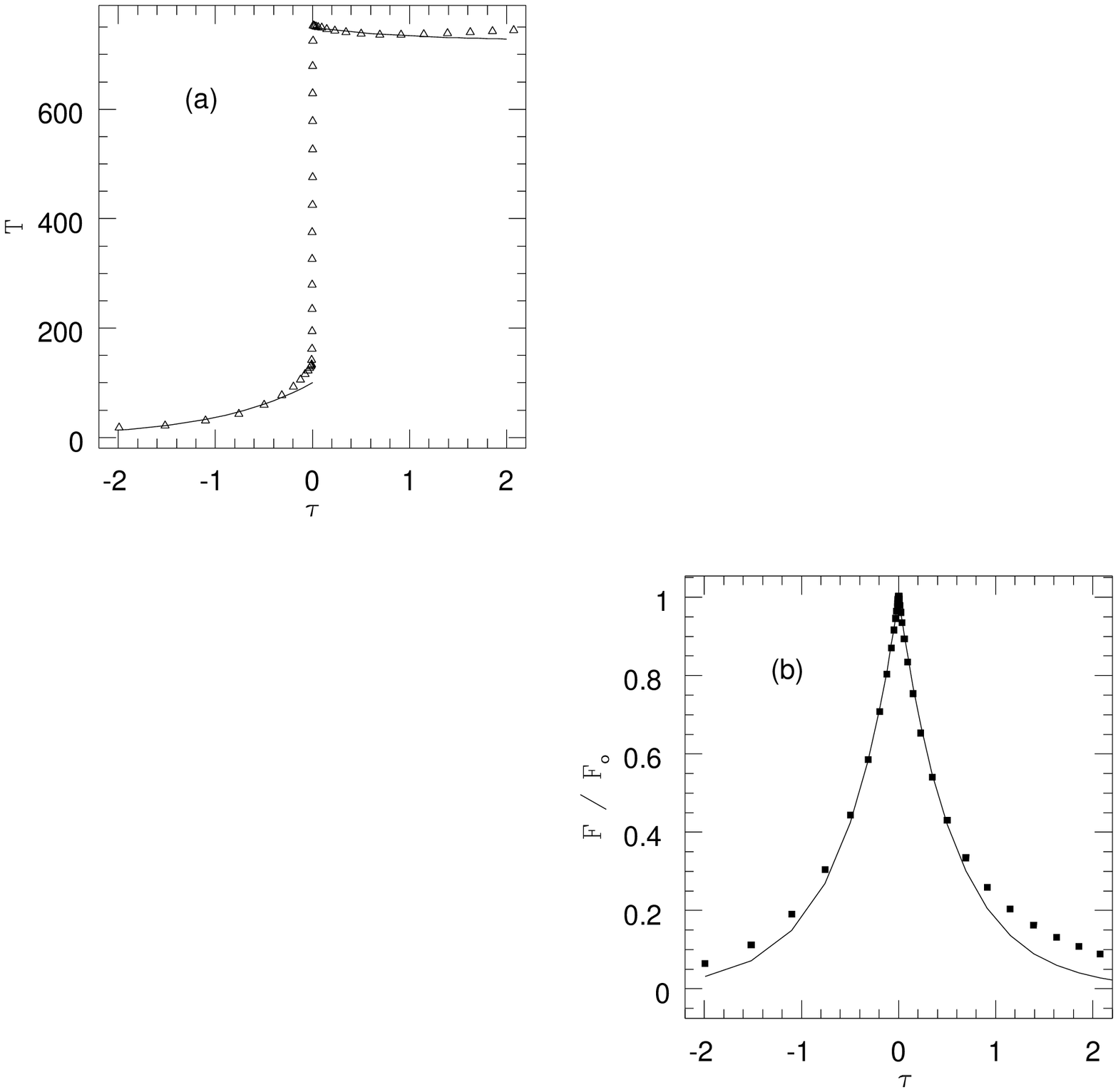}
\caption{\label{fig: flux vs tau}   }
\end{figure}

Figure \ref{fig: flux vs tau}. Gas temperature (a) and radiation flux
(b) as a function of optical depth
for a subcritical shock.  The solid lines represent the analytic solution
from ZR and
the points are the numerical result.

\begin{figure}
\plotone{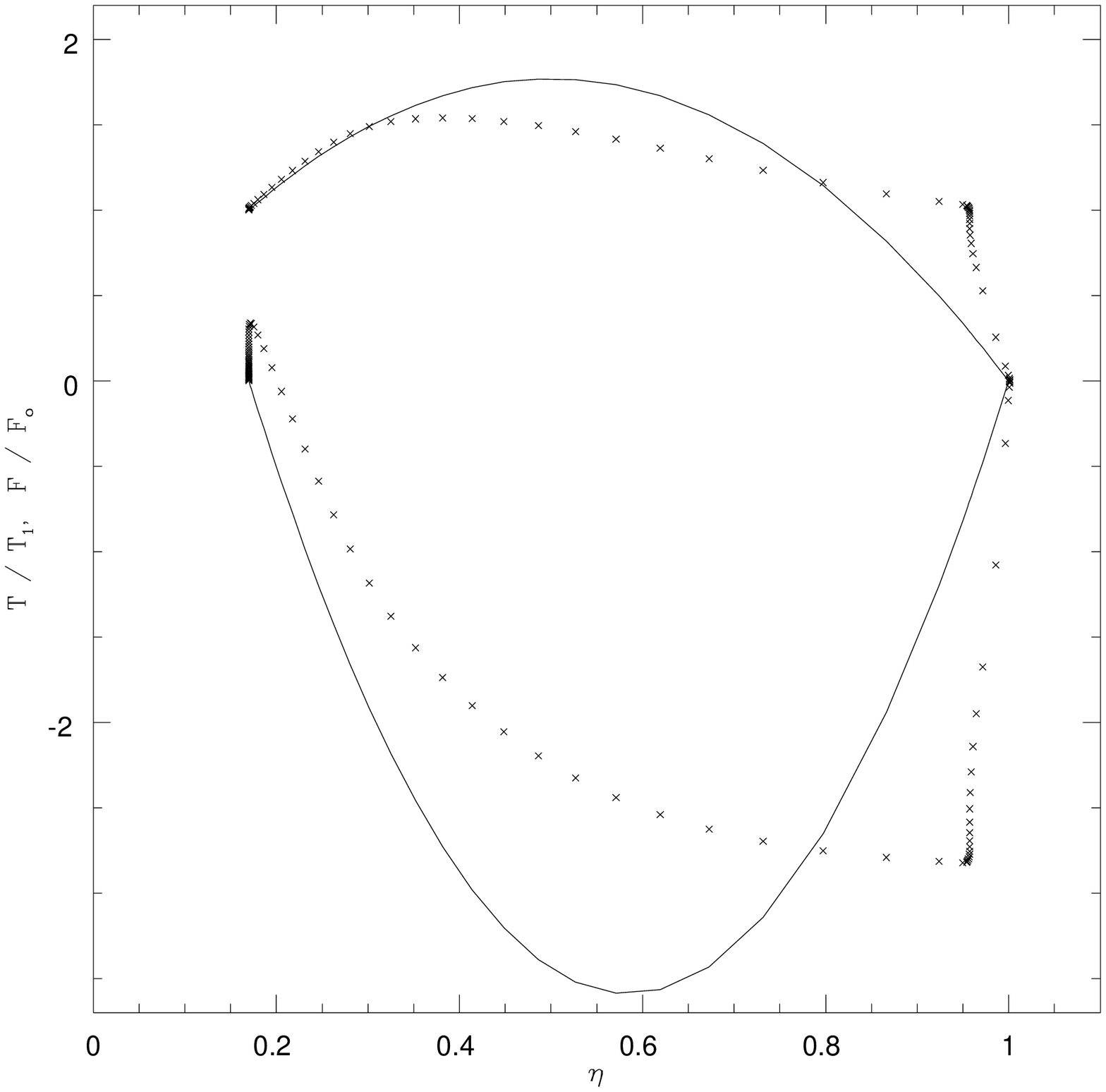}
\caption{\label{fig: supereta}   }
\end{figure}

Figure \ref{fig: supereta}. The gas temperature and radiation as a function
of compression ratio for a supercritical shock wave.

\begin{figure}
\plotone{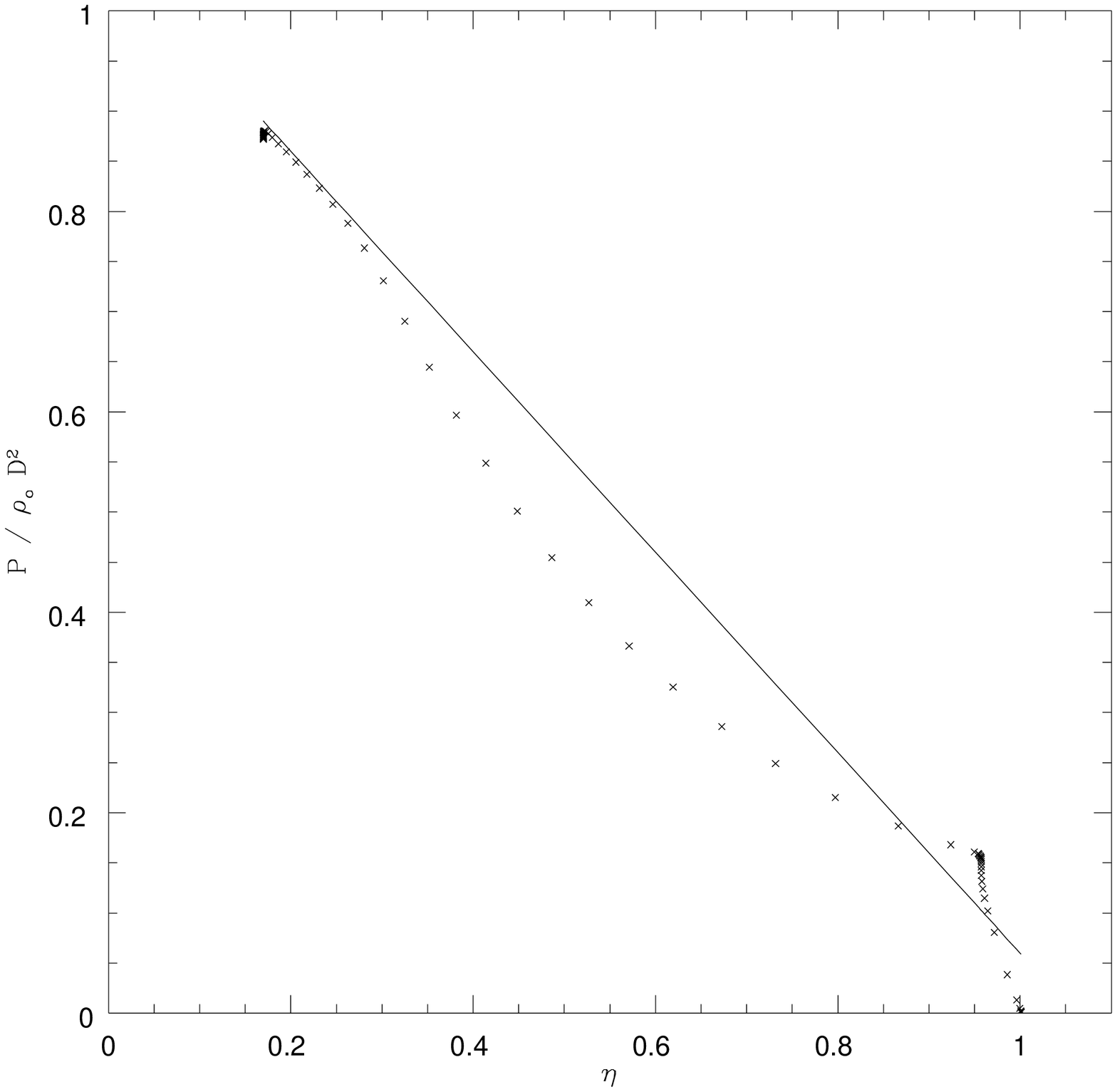}
\caption{\label{fig: superpeta}   }
\end{figure}

Figure \ref{fig: superpeta}. The gas pressure as a function
of compression ratio for a supercritical shock wave.

\begin{figure}
\plotone{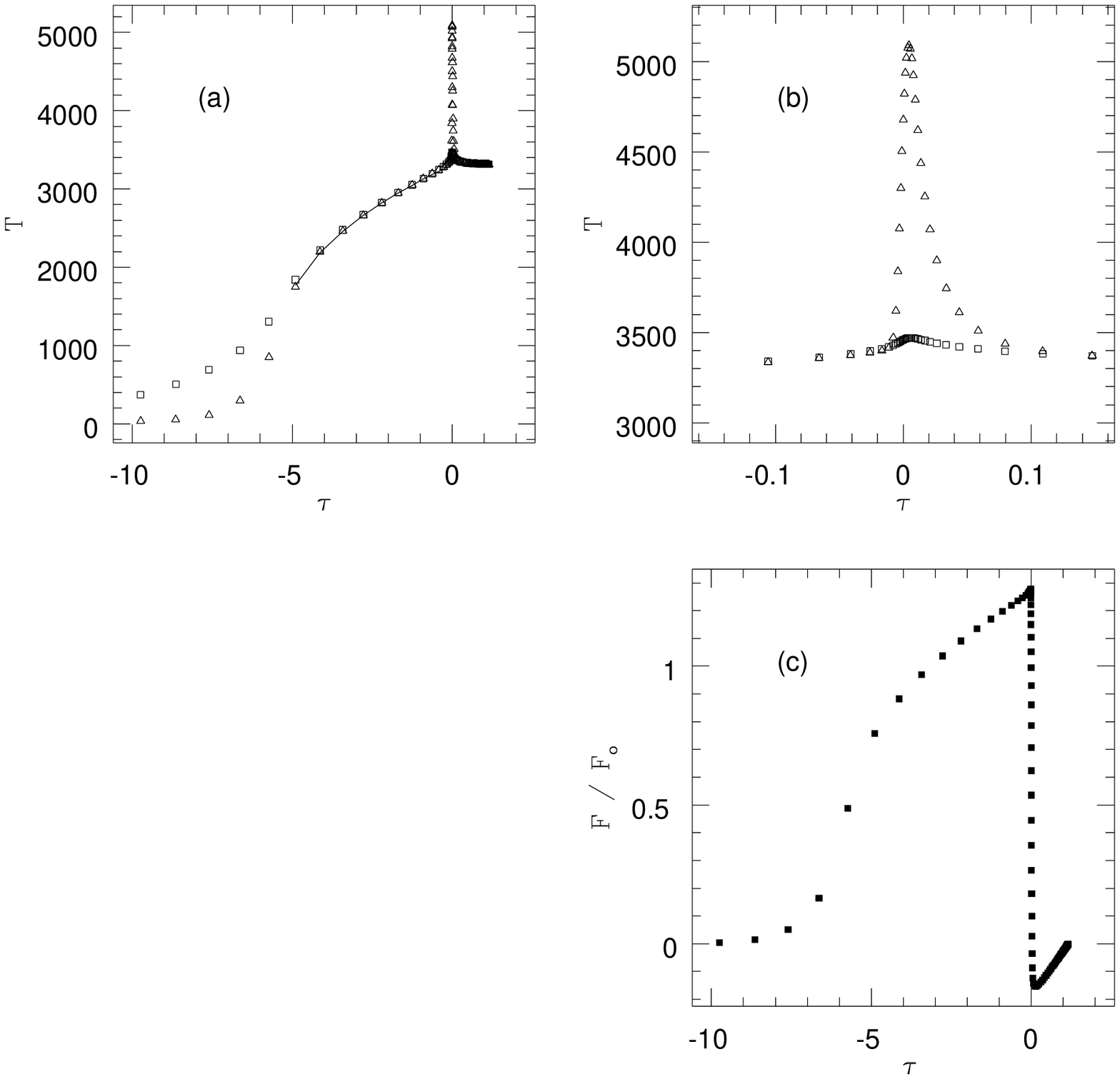}
\caption{\label{fig: super three zone}   }
\end{figure}

Figure \ref{fig: super three zone}. Gas temperature (triangles) and
radiation temperature (squares) as a function of optical depth for a
supercritical shock (a).  The temperature profile is shown for a small
range of $\tau$ in (b).  The radiation flux profile is plotted in (c). 
The solid curves are the analytic approximation for the temperature in the
equilibrium zone. 

\begin{figure}
\plotone{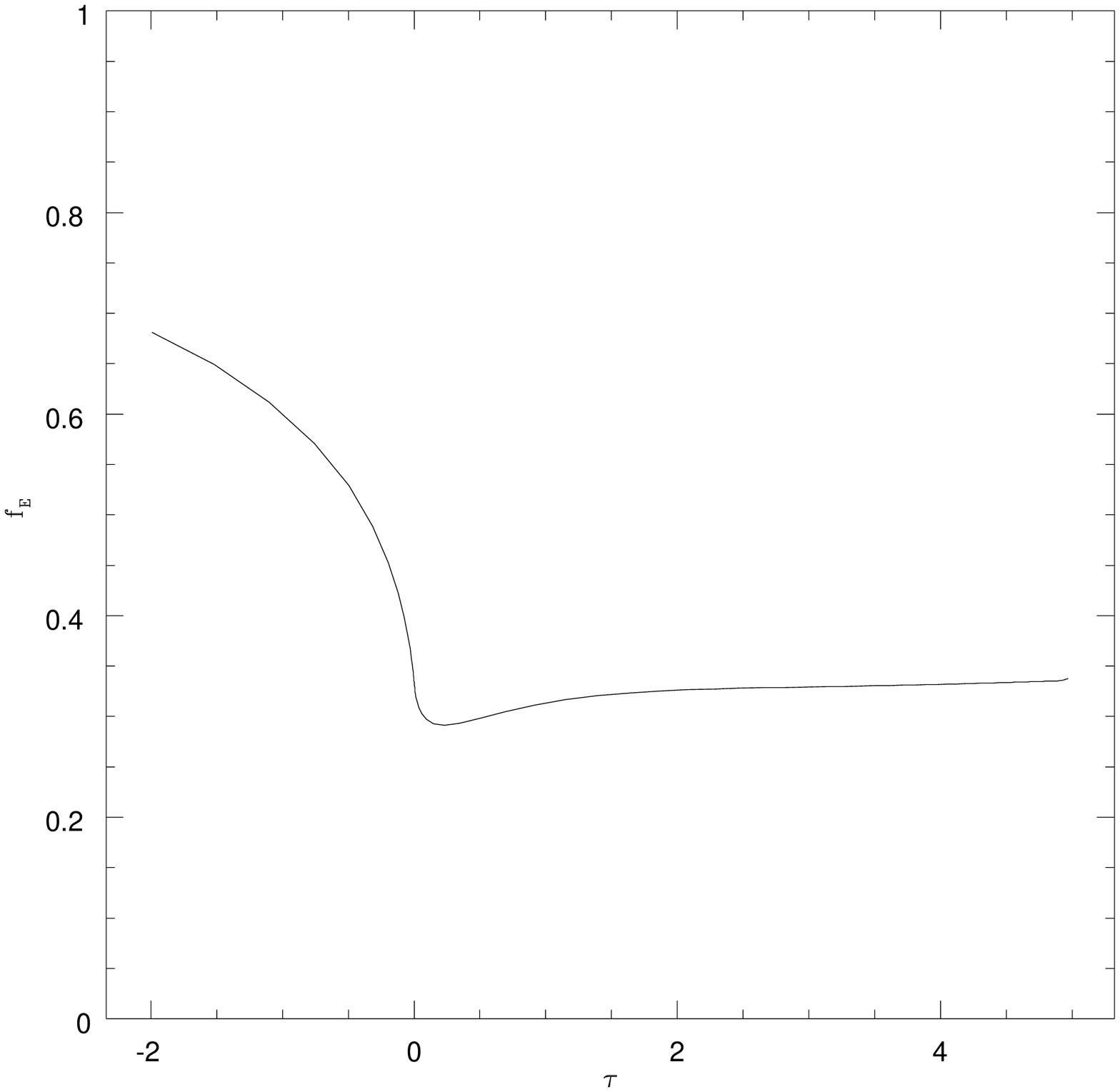}
\caption{\label{fig: sub eddfac}   }
\end{figure}

Figure \ref{fig: sub eddfac}. Eddington factors as a function of
optical depth for a 
subcritical shock.  Note that $f_E < 1/3$ at the shock radius.

\begin{figure}
\plotone{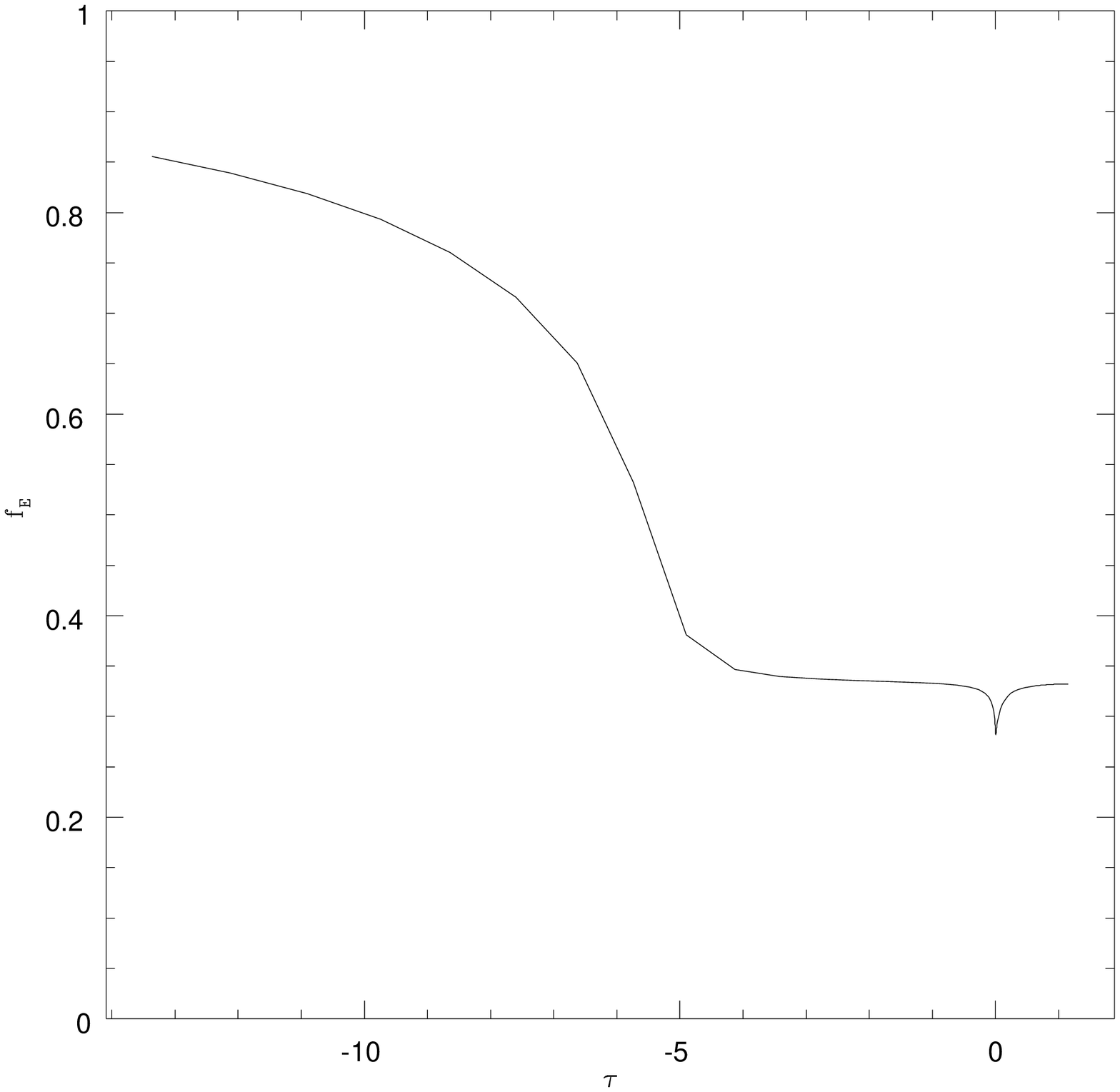}
\caption{\label{fig: super eddfac}   }
\end{figure}

Figure \ref{fig: super eddfac}. Eddington factors as a function of optical
depth for a 
supercritical shock.  Note that $f_E < 1/3$ at the shock radius.

Figure \ref{fig: transport comparison}.  The numerical solutions for full 
transport (crosses), non-equilibrium diffusion (open squares) and
equilibrium diffusion (stars).  The curve represents the analytic
solution.  The piston speeds are, from top to bottom: 12, 9 and 6 km/sec. 

\begin{figure}
\plotone{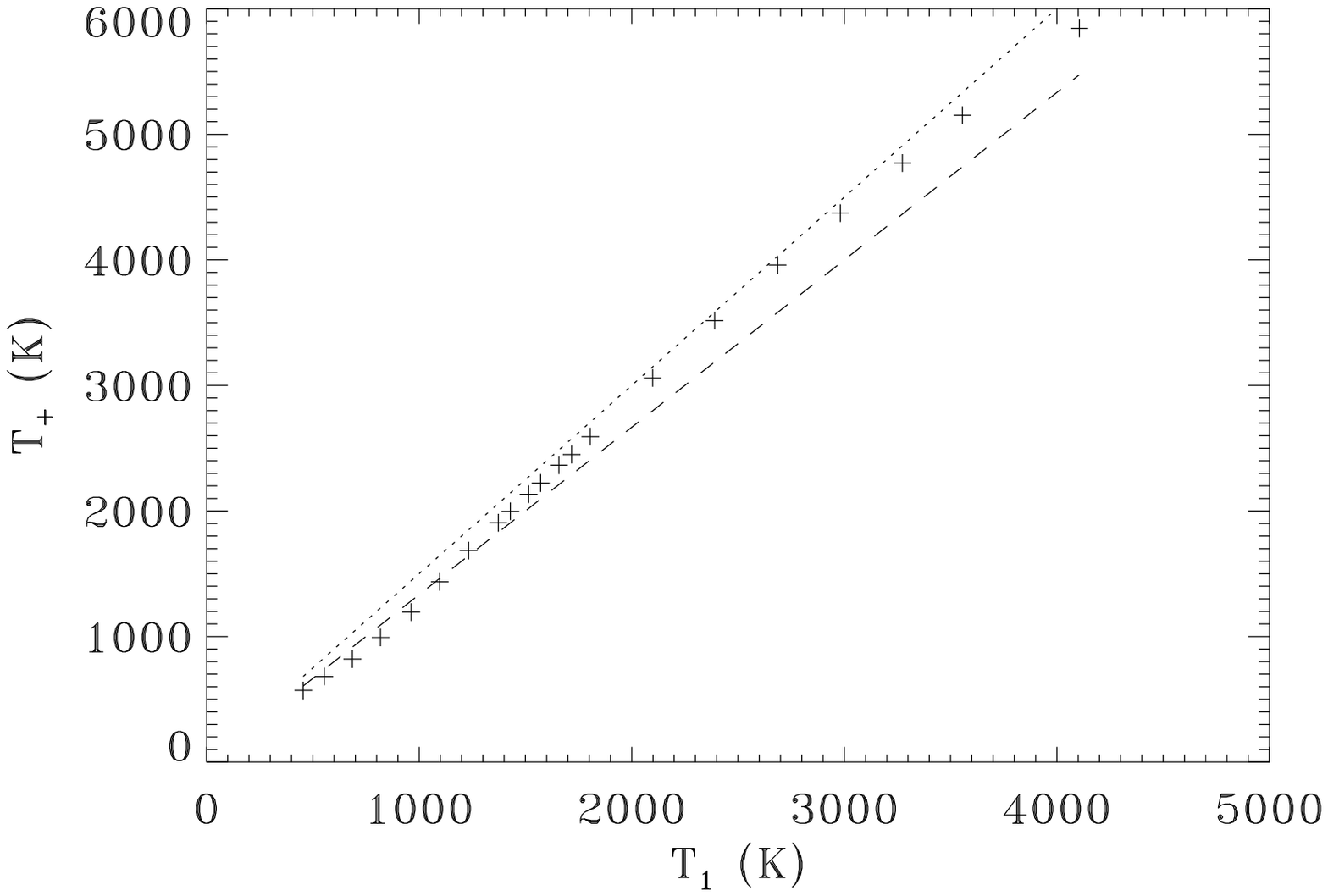}
\caption{\label{fig: tp vs t1}   }
\end{figure}

Figure \ref{fig: tp vs t1}.  The post-shock gas temperature ($T_{+}$) as a 
function of the final temperature of the shocked gas.

\begin{figure}
\plotone{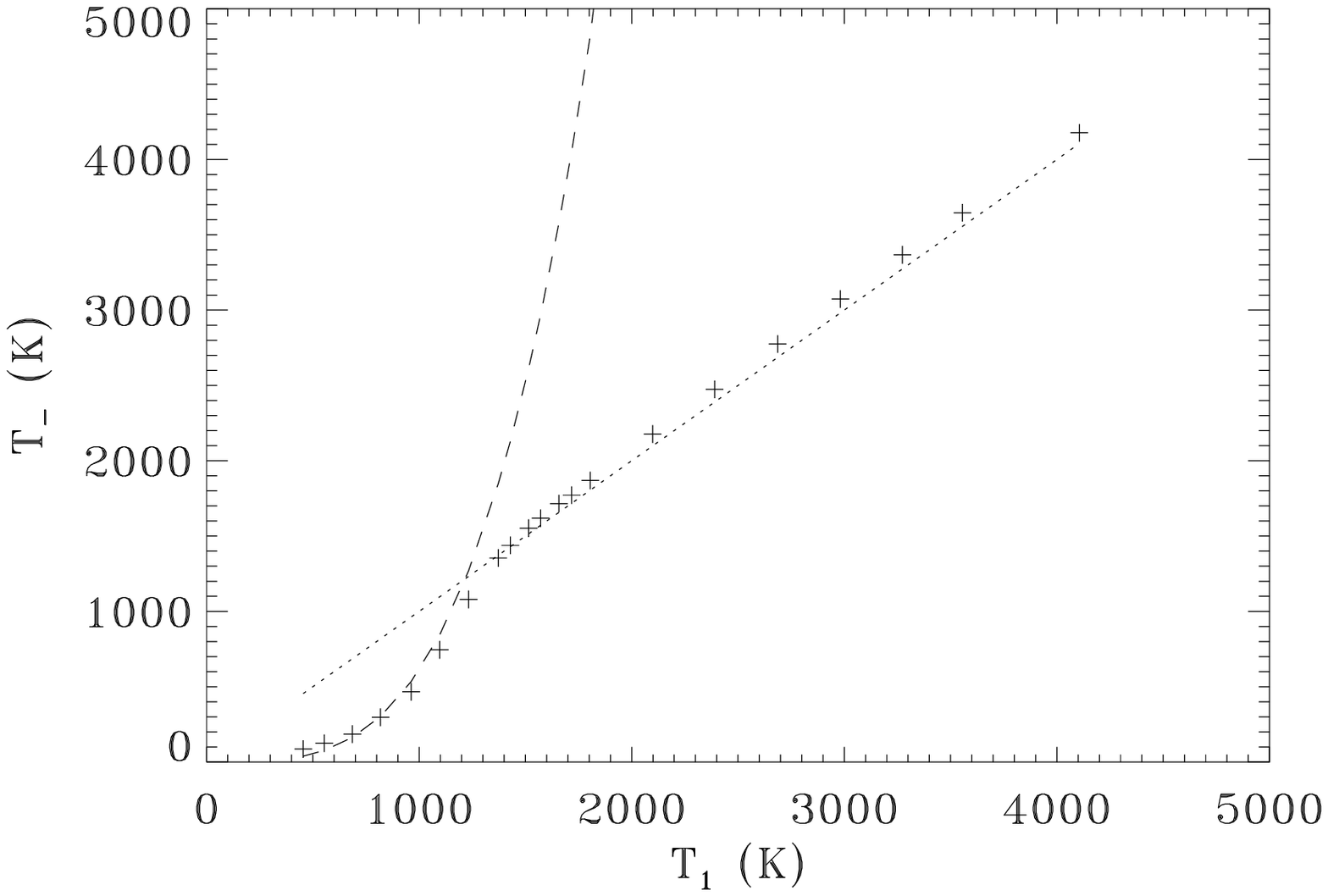}
\caption{\label{fig: tm vs t1}   }
\end{figure}

Figure \ref{fig: tm vs t1}.  The pre-shock gas temperature ($T_{-}$) as a 
function of the final temperature of the shocked gas.

Figure \ref{fig:  color t vs tau}. The gas temperature as a function of
the optical depth. Color coding is described in the appendix.

Figure \ref{fig:  color p vs tau}. The gas pressure as a function of
the optical depth. Color coding is described in the appendix.

Figure \ref{fig:  color f vs tau}. The radiation flux as a function of
the optical depth. Color coding is described in the appendix.

Figure \ref{fig:  color fe vs tau}. The Eddington factors as a function of
the optical depth. Color coding is described in the appendix.

\begin{figure}
\plotone{transport.ps}
\caption{\label{fig: transport comparison}   }
\end{figure}

\begin{figure}
\caption{\label{fig: color t vs tau}   }
\end{figure}

\begin{figure}
\caption{\label{fig: color p vs tau}   }
\end{figure}

\begin{figure}
\caption{\label{fig: color f vs tau}   }
\end{figure}

\begin{figure}
\caption{\label{fig: color fe vs tau}   }
\end{figure}

\end{document}